\pdfoutput=1
\documentclass[12pt,a4paper]{article}

\usepackage{ifthen} % for conditional statements
\newboolean{pdflatex}
\setboolean{pdflatex}{true} % False for eps figures 

\newboolean{articletitles}
\setboolean{articletitles}{true} % False removes titles in references

\newboolean{uprightparticles}
\setboolean{uprightparticles}{false} %True for upright particle symbols

\newboolean{inbibliography}
\setboolean{inbibliography}{false} %True once you enter the bibliography

\usepackage[top=1in, bottom=1.25in, left=1in, right=1in]{geometry}

\usepackage{microtype}
\usepackage{lineno}  % for line numbering during review
\usepackage{xspace} % To avoid problems with missing or double spaces after
\usepackage{caption} %these three command get the figure and table captions automatically small

\usepackage{graphicx}  % to include figures (can also use other packages)
\usepackage{color}
\usepackage{colortbl}
\graphicspath{{./figs/}} % Make Latex search fig subdir for figures

\usepackage{amsmath} % Adds a large collection of math symbols
\usepackage{amssymb}
\usepackage{amsfonts}
\usepackage{upgreek} % Adds in support for greek letters in roman typeset

\newcommand*\patchAmsMathEnvironmentForLineno[1]{%
\expandafter\let\csname old#1\expandafter\endcsname\csname #1\endcsname
\expandafter\let\csname oldend#1\expandafter\endcsname\csname
end#1\endcsname
 \renewenvironment{#1}%
   {\linenomath\csname old#1\endcsname}%
   {\csname oldend#1\endcsname\endlinenomath}%
}
\newcommand*\patchBothAmsMathEnvironmentsForLineno[1]{%
  \patchAmsMathEnvironmentForLineno{#1}%
  \patchAmsMathEnvironmentForLineno{#1*}%
}
\AtBeginDocument{%
\patchBothAmsMathEnvironmentsForLineno{equation}%
\patchBothAmsMathEnvironmentsForLineno{align}%
\patchBothAmsMathEnvironmentsForLineno{flalign}%
\patchBothAmsMathEnvironmentsForLineno{alignat}%
\patchBothAmsMathEnvironmentsForLineno{gather}%
\patchBothAmsMathEnvironmentsForLineno{multline}%
\patchBothAmsMathEnvironmentsForLineno{eqnarray}%
}

\usepackage{hyperref}    % Hyperlinks in references
\usepackage[all]{hypcap} % Internal hyperlinks to floats.

\usepackage{xspace} 
\usepackage{upgreek}

\def\lhcb {\mbox{LHCb}\xspace}

\def\babar  {\mbox{BaBar}\xspace}

\def\MagUp {\mbox{\em Mag\kern -0.05em Up}\xspace}

\ifthenelse{\boolean{uprightparticles}}%
{

 \def\Ppi         {\ensuremath{\uppi}\xspace}

 \def\Ptau        {\ensuremath{\uptau}\xspace}

 \def\PDelta      {\ensuremath{\Delta}\xspace}                 
 \def\PXi      {\ensuremath{\Xi}\xspace}                 
 \def\PLambda      {\ensuremath{\Lambda}\xspace}                 
 \def\PSigma      {\ensuremath{\Sigma}\xspace}                 
 \def\POmega      {\ensuremath{\Omega}\xspace}                 
 \def\PUpsilon      {\ensuremath{\Upsilon}\xspace}                 
 
 \def\PB      {\ensuremath{\mathrm{B}}\xspace}                 
                  
 \def\PD      {\ensuremath{\mathrm{D}}\xspace}

 \def\PK      {\ensuremath{\mathrm{K}}\xspace}

 \def\Pb      {\ensuremath{\mathrm{b}}\xspace}

 \def\Pi      {\ensuremath{\mathrm{i}}\xspace}

 \def\Ps      {\ensuremath{\mathrm{s}}\xspace}

}
{

 \def\Ppi         {\ensuremath{\pi}\xspace}

 \def\Ptau        {\ensuremath{\tau}\xspace}

 \mathchardef\PDelta="7101
 \mathchardef\PXi="7104
 \mathchardef\PLambda="7103
 \mathchardef\PSigma="7106
 \mathchardef\POmega="710A
 \mathchardef\PUpsilon="7107
                  
 \def\PB      {\ensuremath{B}\xspace}                 
                  
 \def\PD      {\ensuremath{D}\xspace}

 \def\PK      {\ensuremath{K}\xspace}

 \def\Pb      {\ensuremath{b}\xspace}

 \def\Pi      {\ensuremath{i}\xspace}

 \def\Ps      {\ensuremath{s}\xspace}

}

\makeatletter
\ifcase \@ptsize \relax% 10pt
  \newcommand{\miniscule}{\@setfontsize\miniscule{4}{5}}% \tiny: 5/6
\or% 11pt
  \newcommand{\miniscule}{\@setfontsize\miniscule{5}{6}}% \tiny: 6/7
\or% 12pt
  \newcommand{\miniscule}{\@setfontsize\miniscule{5}{6}}% \tiny: 6/7
\fi
\makeatother

\DeclareRobustCommand{\optbar}[1]{\shortstack{{\miniscule (\rule[.5ex]{1.25em}{.18mm})}
  \\ [-.7ex] $#1$}}

\def\taum       {{\ensuremath{\Ptau^-}}\xspace}
\def\tautau     {{\ensuremath{\Ptau^+\Ptau^-}}\xspace}

\def\squark    {{\ensuremath{\Ps}}\xspace}

\def\bquark    {{\ensuremath{\Pb}}\xspace}

\def\pion   {{\ensuremath{\Ppi}}\xspace}

\def\pip    {{\ensuremath{\pion^+}}\xspace}
\def\pim    {{\ensuremath{\pion^-}}\xspace}

\def\kaon    {{\ensuremath{\PK}}\xspace}
  \def\Kbar    {{\kern 0.2em\overline{\kern -0.2em \PK}{}}\xspace}

\def\KorKbar    {\kern 0.18em\optbar{\kern -0.18em K}{}\xspace}
\def\Kz      {{\ensuremath{\kaon^0}}\xspace}

\def\Kp      {{\ensuremath{\kaon^+}}\xspace}
\def\Km      {{\ensuremath{\kaon^-}}\xspace}

  \def\Dbar    {{\kern 0.2em\overline{\kern -0.2em \PD}{}}\xspace}
\def\D       {{\ensuremath{\PD}}\xspace}

\def\DorDbar    {\kern 0.18em\optbar{\kern -0.18em D}{}\xspace}

\def\Dp      {{\ensuremath{\D^+}}\xspace}
\def\Dm      {{\ensuremath{\D^-}}\xspace}

\def\Ds      {{\ensuremath{\D^+_\squark}}\xspace}
\def\Dsp     {{\ensuremath{\D^+_\squark}}\xspace}
\def\Dsm     {{\ensuremath{\D^-_\squark}}\xspace}

\def\B       {{\ensuremath{\PB}}\xspace}
\def\Bbar    {{\ensuremath{\kern 0.18em\overline{\kern -0.18em \PB}{}}}\xspace}

\def\BorBbar    {\kern 0.18em\optbar{\kern -0.18em B}{}\xspace}
\def\Bz      {{\ensuremath{\B^0}}\xspace}

\def\Bd      {{\ensuremath{\B^0}}\xspace}
\def\Bs      {{\ensuremath{\B^0_\squark}}\xspace}

  \def\Y#1S{\ensuremath{\PUpsilon{(#1S)}}\xspace}% no space before {...}!

\def\Lbar        {{\ensuremath{\kern 0.1em\overline{\kern -0.1em\PLambda}}}\xspace}
\def\LorLbar    {\kern 0.18em\optbar{\kern -0.18em \PLambda}{}\xspace}

\def\BF         {{\ensuremath{\mathcal{B}}}\xspace}

\def\BR         {\BF}
\def\to                 {\ensuremath{\rightarrow}\xspace}

\def\eps   {{\ensuremath{\varepsilon}}\xspace}

\def\AT#1     {\ensuremath{A_{\mathrm{T}}^{#1}}\xspace}           % 2

\def\C#1      {\ensuremath{\mathcal{C}_{#1}}\xspace}                       % 9
\def\Cp#1     {\ensuremath{\mathcal{C}_{#1}^{'}}\xspace}                    % 7
\def\Ceff#1   {\ensuremath{\mathcal{C}_{#1}^{\mathrm{(eff)}}}\xspace}        % 9  
\def\Cpeff#1  {\ensuremath{\mathcal{C}_{#1}^{'\mathrm{(eff)}}}\xspace}       % 7
\def\Ope#1    {\ensuremath{\mathcal{O}_{#1}}\xspace}                       % 2
\def\Opep#1   {\ensuremath{\mathcal{O}_{#1}^{'}}\xspace}                    % 7

\newcommand{\tev}{\ifthenelse{\boolean{inbibliography}}{\ensuremath{~T\kern -0.05em eV}}{\ensuremath{\mathrm{\,Te\kern -0.1em V}}}\xspace}
\newcommand{\gev}{\ensuremath{\mathrm{\,Ge\kern -0.1em V}}\xspace}
\newcommand{\mev}{\ensuremath{\mathrm{\,Me\kern -0.1em V}}\xspace}
\newcommand{\kev}{\ensuremath{\mathrm{\,ke\kern -0.1em V}}\xspace}
\newcommand{\ev}{\ensuremath{\mathrm{\,e\kern -0.1em V}}\xspace}
\newcommand{\gevc}{\ensuremath{{\mathrm{\,Ge\kern -0.1em V\!/}c}}\xspace}
\newcommand{\mevc}{\ensuremath{{\mathrm{\,Me\kern -0.1em V\!/}c}}\xspace}
\newcommand{\gevcc}{\ensuremath{{\mathrm{\,Ge\kern -0.1em V\!/}c^2}}\xspace}
\newcommand{\gevgevcccc}{\ensuremath{{\mathrm{\,Ge\kern -0.1em V^2\!/}c^4}}\xspace}
\newcommand{\mevcc}{\ensuremath{{\mathrm{\,Me\kern -0.1em V\!/}c^2}}\xspace}

\def\invfb   {\ensuremath{\mbox{\,fb}^{-1}}\xspace}

\newcommand{\stat}{\ensuremath{\mathrm{\,(stat)}}\xspace}
\newcommand{\syst}{\ensuremath{\mathrm{\,(syst)}}\xspace}

\def\gsim{{~\raise.15em\hbox{$>$}\kern-.85em
          \lower.35em\hbox{$\sim$}~}\xspace}
\def\lsim{{~\raise.15em\hbox{$<$}\kern-.85em
          \lower.35em\hbox{$\sim$}~}\xspace}

\newcommand{\Real}{\ensuremath{\mathcal{R}e}\xspace}
\newcommand{\Imag}{\ensuremath{\mathcal{I}m}\xspace}

\def\pt         {\mbox{$p_{\mathrm{ T}}$}\xspace}

\def\evtgen     {\mbox{\textsc{EvtGen}}\xspace}

\def\geant      {\mbox{\textsc{Geant4}}\xspace}

\def\photos     {\mbox{\textsc{Photos}}\xspace}

\def\pythia     {\mbox{\textsc{Pythia}}\xspace}

\def\tell1  {TELL1\xspace}
\def\ukl1   {UKL1\xspace}

\newcommand{\eg}{\mbox{\itshape e.g.}\xspace}
\newcommand{\ie}{\mbox{\itshape i.e.}\xspace}

\usepackage{cite} % Allows for ranges in citations
\usepackage{mciteplus}

\makeatletter
\g@addto@macro\bfseries{\boldmath}
\makeatother

\begin{document}

%%%%%%%%%%%%%%
%%%%% Title     %%%%%
%%%%%%%%%%%%%%
\renewcommand{\thefootnote}{\fnsymbol{footnote}}
\setcounter{footnote}{1}

%%%%%%%%%%%%%%%%%%
%%%%%  TITLE PAGE  %%%%%
%%%%%%%%%%%%%%%%%%
\begin{titlepage}
\pagenumbering{roman}

% Header ---------------------------------------------------
\vspace*{-1.5cm}
\centerline{\large EUROPEAN ORGANIZATION FOR NUCLEAR RESEARCH (CERN)}
\vspace*{1.5cm}
\noindent
\begin{tabular*}{\linewidth}{lc@{\extracolsep{\fill}}r@{\extracolsep{0pt}}}
\ifthenelse{\boolean{pdflatex}}% Logo format choice
{\vspace*{-2.7cm}\mbox{\!\!\!\includegraphics[width=.14\textwidth]{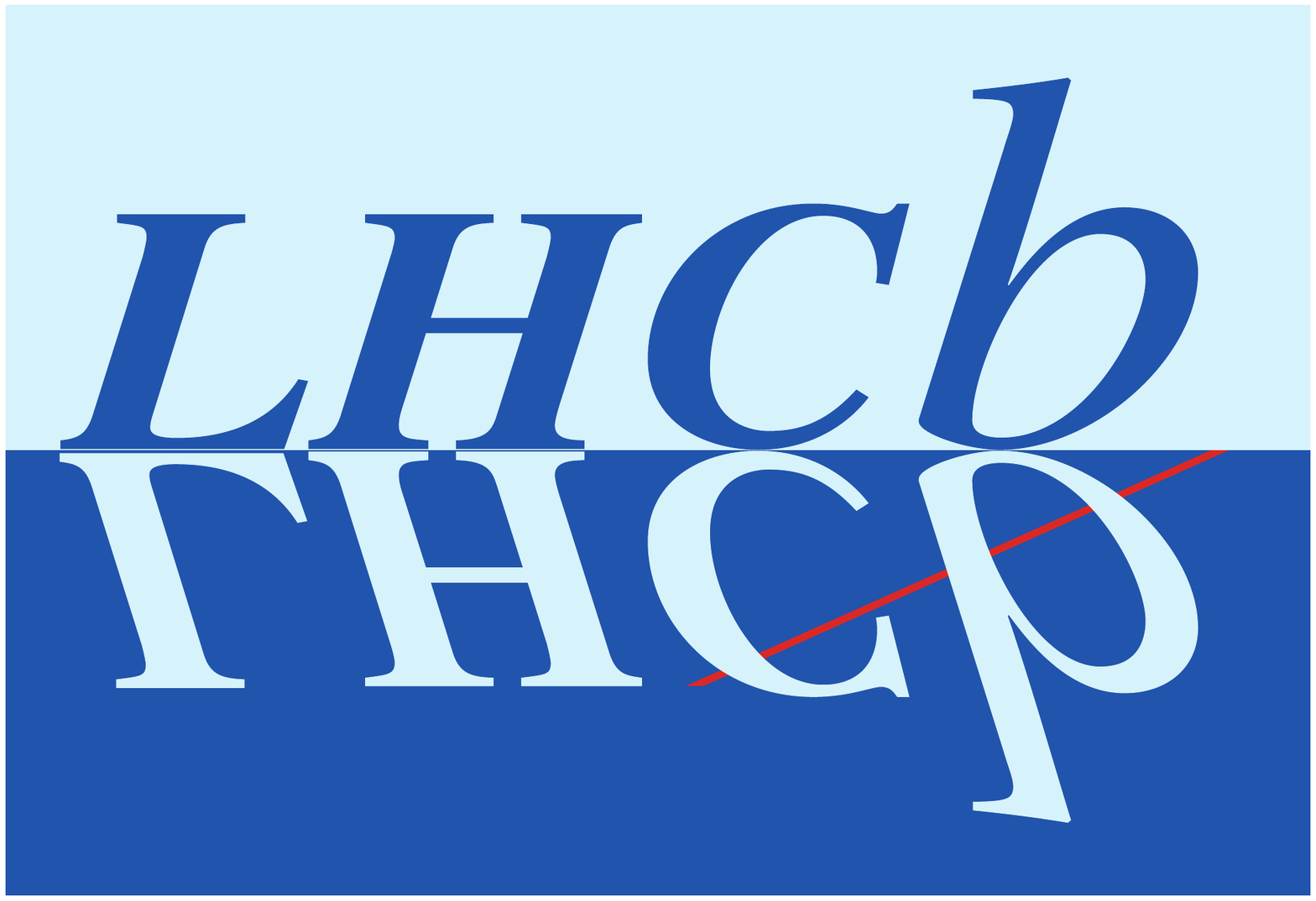}} & &}%
%{\vspace*{-1.2cm}\mbox{\!\!\!\includegraphics[width=.14\textwidth]{lhcb-logo.pdf}} & &}%
{\vspace*{-1.2cm}\mbox{\!\!\!\includegraphics[width=.12\textwidth]{lhcb-logo.eps}} & &}%
\\
 & & CERN-EP-2017-034 \\  % ID 
 & & LHCb-PAPER-2017-003 \\  % ID 
 & & \today \\ % Date
\end{tabular*}

\vspace*{4.0cm}

% Title --------------------------------------------------
{\normalfont\bfseries\boldmath\huge
\begin{center}
Search for the decays $\Bs\to\tautau$ and $\Bd\to\tautau$
\end{center}
}

\vspace*{2.0cm}

% Authors -------------------------------------------------
\begin{center}
The LHCb collaboration\footnote{Authors are listed at the end of this Letter.}
\end{center}

\vspace{\fill}

% Abstract -----------------------------------------------
\begin{abstract}
\noindent
A search for the rare decays $\Bs\to\tautau$ and $\Bd\to\tautau$ is performed using proton--proton collision data collected with the LHCb detector.
The data sample corresponds to an integrated luminosity of 3\invfb collected in 2011 and 2012.
The $\tau$ leptons are reconstructed through the decay $\taum\to\pim\pip\pim\nu_{\tau}$.
Assuming no contribution from $\Bd\to\tautau$ decays, an upper limit is set on the branching fraction $\BR(\Bs\to\tautau) < 6.8\times 10^{-3}$ at 95\% confidence level.
If instead no contribution from $\Bs\to\tautau$ decays is assumed, the limit is $\BR(\Bd\to\tautau)< 2.1 \times 10^{-3}$ at 95\% confidence level.
These results correspond to the first direct limit on $\BR(\Bs\to\tautau)$ and the world's best limit on $\BR(\Bd\to\tautau)$.
\end{abstract}

\vspace*{2.0cm}

\begin{center}
Published in Phys.~Rev.~Lett.~118 (2017) 251802
\end{center}

\vspace{\fill}

{\footnotesize 
\centerline{\copyright~CERN on behalf of the \lhcb collaboration, licence \href{http://creativecommons.org/licenses/by/4.0/}{CC-BY-4.0}.}}
\vspace*{2mm}

\end{titlepage}

%%%%%%%%%%%%%%%%%%%%%%%%%%%%%%%%
%%%%%  EOD OF TITLE PAGE  %%%%%%
%%%%%%%%%%%%%%%%%%%%%%%%%%%%%%%%

%  empty page follows the title page ----
\newpage
\setcounter{page}{2}
\mbox{~}

\cleardoublepage

\renewcommand{\thefootnote}{\arabic{footnote}}
\setcounter{footnote}{0}

%%%%%%%%%%%%%%%
%%%%% Main text %%%%%
%%%%%%%%%%%%%%%

\pagestyle{plain} % restore page numbers for the main text
\setcounter{page}{1}
\pagenumbering{arabic}

%% Uncomment during review phase. 
%% Comment before a final submission.
%\linenumbers

%%%%% Main Body
%!TEX root = main.tex

%%%%%%%%%%%%%%%%%%%%%%%%%%%%%%%%%%%%%%%%%%%%%%%%%%
%%% Introduction
\noindent
Processes where a $B$ meson decays into a pair of oppositely charged leptons are powerful probes in the search for physics beyond the Standard Model (SM).
Recently, the first observation of the $\Bs\to\mu^+\mu^-$ decay was made \cite{LHCb-PAPER-2014-049, LHCb-PAPER-2017-001} (the inclusion of charge-conjugate processes is implied throughout this Letter).
Its measured branching fraction ($\BR$) is compatible with the SM prediction \cite{Bobeth:2013uxa} and imposes stringent constraints on theories beyond the SM.
Complementing this result with searches for the tauonic modes \mbox{$B\to\tautau$,} where $B$ can be either a \Bd or a \Bs meson, is of great interest in view of the recent hints of lepton flavour non-universality obtained by several experiments.
In particular the measurements of $R(D^{(*)}) = \frac{\BR(\Bd\to D^{(*)}\tau^+\nu_{\tau})}{\BR(\Bd\to D^{(*)}\ell^+\nu_{\ell})}$, where $\ell^+$ represents either a muon, an electron or both, are found to be larger than the SM prediction by 3.9 standard deviations ($\sigma$) \cite{HFAG}, and the measurement of $R_K= \frac{\BR (B^+ \to K^+ \mu^+ \mu^-)} {\BR (B^+ \to K^+ e^+ e^-)}$ is 2.6$\sigma$ lower than the SM prediction \cite{LHCb-PAPER-2014-024}.
Possible explanations for these and other \cite{LHCb-PAPER-2015-051} deviations from their SM expectations include leptoquarks, $W'/Z'$ bosons and two-Higgs-doublet models (see \eg Refs.~\cite{Crivellin:2015lwa, Becirevic:2016yqi}).
In these models, the $B\to\tautau$ branching fractions could be enhanced with respect to the SM predictions, $\BR(\Bd\to\tautau) = (2.22 \pm 0.19)\times 10^{-8}$ and $\BR(\Bs\to\tautau) = (7.73 \pm 0.49)\times 10^{-7}$ \cite{Bobeth:2013uxa}, by several orders of magnitude \cite{Dighe:2012df, Alonso:2015sja,Cline:2015lqp,Becirevic:2016yqi,Becirevic:2016oho}.
All minimal-flavour-violating models predict the same enhancement of $\BR(\Bs\to\tautau)$ over $\BR(\Bd\to\tautau)$  as in the SM.

The experimental search for $B\to\tautau$ decays is complicated by the presence of at least two undetected neutrinos, originating from the decay of the $\tau$ leptons.
The \babar collaboration has searched for the $\Bd\to\tautau$ mode \cite{Aubert:2005qw} and published an upper limit $\BR(\Bz\to\tautau)<4.10 \times 10^{-3}$ at 90\% confidence level (CL). 
There are currently no experimental results for the $\Bs\to\tautau$ mode, though its branching fraction can be indirectly constrained to be less than 3\% at 90\% CL \cite{Grossman:1996qj,Dighe:2010nj,Bobeth:2011st}.

In this Letter, the first search for the rare decay $\Bs\to\tautau$ is presented, along with a search for the $\Bd\to\tautau$ decay.
The analysis is performed with proton--proton collision data corresponding to integrated luminosities of 1.0\invfb and 2.0\invfb recorded with the LHCb detector at centre-of-mass energies of 7 and 8\,\tev, respectively.
The $\tau$ leptons are reconstructed through the decay $\taum\to\pim\pip\pim\nu_{\tau}$, which proceeds predominantly through the decay chain $\taum \to a_1(1260)^-\nu_{\tau}$, $a_1(1260)^-\to \rho(770)^0\pi^-$ \cite{Schael:2005am}.
The branching fraction $\BR(\tau^-\to \pi^-\pi^+\pi^-\nu_{\tau})$ is $(9.31 \pm 0.05)\%$ \cite{PDG2016}.
Due to the final-state neutrinos the \tautau mass provides only a weak discrimination between signal and background, and cannot be used as a way to distinguish $\Bs$ from $\Bd$ decays.
The number of signal candidates is obtained from a fit to the output of a multivariate classifier that uses a range of kinematic and topological variables as input.
Data-driven methods are used to determine  signal and background models.
The observed signal yield is converted into a branching fraction using as a normalisation channel the decay $\Bd\to\Dm\Dsp$ \cite{Zupanc:2007pu, Aubert:2006nm}, with $\Dm\to\Kp\pim\pim$ and $\Dsp\to\Km\Kp\pip$.

%%%%%%%%%%%%%%%%%%%%%%%%%%%%%%%%%%%%%%%%%%%%%%%%%%
%%% Detector and simulation
The LHCb detector, described in detail in Refs.~\cite{Alves:2008zz,LHCb-DP-2014-002}, is a single-arm forward spectrometer covering the pseudorapidity range $2 < \eta < 5$.
The online event selection is performed by a trigger~\cite{LHCb-DP-2012-004}, which consists of a hardware stage, based on information from the calorimeter and muon systems, followed by a software stage, which applies a full event reconstruction.
The hardware trigger stage requires events to have a muon with high transverse momentum (\pt) with respect to the beam line or a hadron, photon or electron with high transverse energy in the calorimeters.
For hadrons, the transverse energy threshold is around 3.5\gev, depending on the data-taking conditions.
The software trigger requires  a two-, three- or four-track secondary vertex with a significant displacement from the  primary $pp$ interaction vertices (PVs).
A multivariate classifier~\cite{BBDT} is used for the identification of secondary vertices that are significantly displaced from the PVs, and are consistent with the decay of a \bquark hadron.
At least one charged particle must have $\pt > 1.7\gevc$ and be inconsistent with originating from any PV.

Simulated data are used to optimise the selection, obtain the signal model for the fit and determine the selection efficiencies.
In the simulation, $pp$ collisions are generated using \pythia~\cite{Sjostrand:2006za,*Sjostrand:2007gs} with a specific \lhcb configuration~\cite{LHCb-PROC-2010-056}.
Decays of hadrons are described by \evtgen~\cite{Lange:2001uf}, in which final-state radiation is generated using \photos~\cite{Golonka:2005pn}. 
The interaction of the generated particles with the detector, and its response, are implemented using the \geant toolkit \cite{Allison:2006ve, *Agostinelli:2002hh} as described in Ref.~\cite{LHCb-PROC-2011-006}.
The $\taum\to\pim\pip\pim\nu_{\tau}$ decays are generated using the resonance chiral Lagrangian model \cite{Nugent:2013hxa} with a tuning based on the \babar results for the $\taum\to\pim\pip\pim\nu_{\tau}$ decays \cite{Nugent:2013ij}, implemented in the \textsc{Tauola} generator \cite{Davidson:2010rw}.

%%%%%%%%%%%%%%%%%%%%%%%%%%%%%%%%%%%%%%%%%%%%%%%%%%
%%% Event selection
In the offline selection of the candidate signal and normalisation decays, requirements on the particle identification (PID) \cite{LHCb-DP-2012-003}, track quality and the impact parameter with respect to any PV are imposed on all charged final-state particles.
Three charged tracks, identified as pions for the $B\to\tautau$ decays, and pions or kaons for the  $\Bd\to\Dm\Dsp$ decays, forming a good-quality vertex are combined to make intermediate $\tau$, \Dp and \Ds candidates. 
The kinematic properties of these candidates, like momenta and masses, are calculated from the three-track combinations.
The flight directions of the $\tau$, \Dp and \Ds candidates are estimated from their calculated momentum vectors.
For the $\tau$ candidates this is a biased estimate due to the missing neutrinos. 
In turn, $B$-meson candidates are reconstructed from two oppositely charged $\tau$ or from $\Dm$ and $\Dsp$ candidates with decay vertices well separated from the PVs. 
The $B$-meson candidates are required to have $\pt > 2\gevc$, at least one $\tau$, \Dp and \Ds candidate with $\pt > 4\gevc$ and at least one pion or kaon with $\pt > 2\gevc$. 
No further selection requirements are imposed on the normalisation mode.

\begin{figure}[!t]
\center
\includegraphics[width=0.7\textwidth]{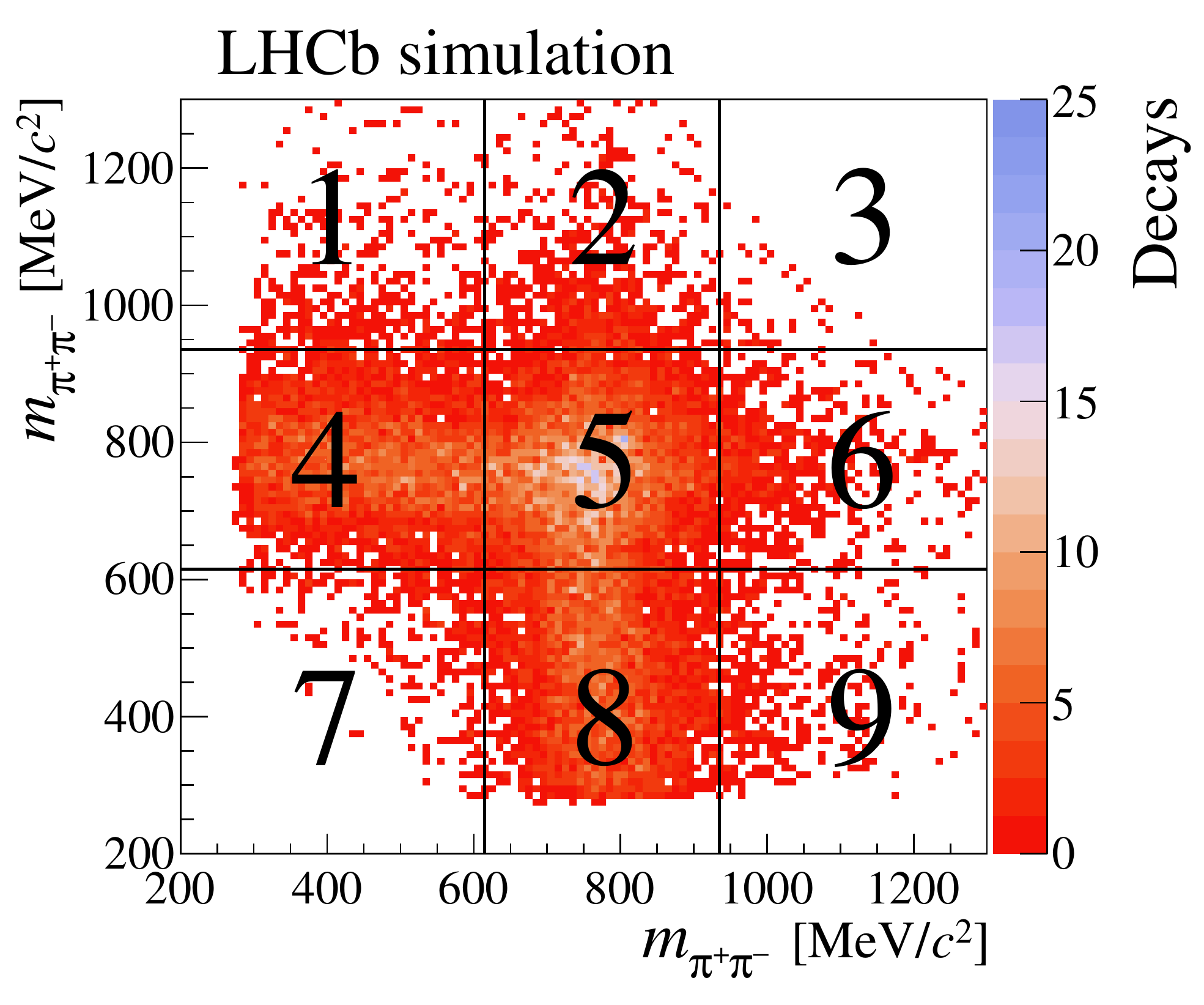}
\caption{Two-dimensional distribution of the invariant masses $m_{\pip\pim}$ of the two oppositely charged two-pion combinations for simulated $\Bs\to\tautau$ candidates.
The distribution is symmetric by construction.
The vertical and horizontal lines illustrate the sector boundaries.}
\label{Fig:Dalitz_Dialpad}
\end{figure}

For each $\tau$ candidate, the two-dimensional distribution of the invariant masses $m_{\pip\pim}$ of the two oppositely charged two-pion combinations is divided into nine sectors, as illustrated in Fig.~\ref{Fig:Dalitz_Dialpad}.
Exploiting the intermediate $\rho(770)^0$ resonance of the $\tau$ decays, these sectors are used to define three regions.
The signal region consists of $B$ candidates with both $\tau$ candidates in sector 5, and is used to determine the signal yield.
The signal-depleted region, composed of $B$ candidates having at least one $\tau$ candidate in sectors 1, 3, 7 or 9, provides a sample used when optimising the selection.
The control region corresponds to $B$ candidates with one $\tau$ candidate in sectors 4, 5 or 8 and the other in sectors 4 or 8, and provides the background model.

%%%%%%%%%%%%%%%%%%%%
%%%%%%%%%%%%%%%%%%%%

For the $\B\to\tautau$ modes, further requirements are imposed on two types of isolation variables that are able to discriminate signal from background from partially reconstructed decays with additional charged or neutral particles.
The first class of isolation variables, based on the decision of a multivariate classifier trained on simulated signal and other $b$-hadron decays, discriminates against processes containing additional charged tracks that either make a good-quality vertex with any selected pion or $\tau$ candidate, or belong to the same $b$-hadron decay as the selected pion candidates.
The second class of isolation variables is based on calorimeter activity due to neutral particles in a cone, defined in terms of the pseudorapidity and polar angle, centred on the $B$ candidate momentum.

In addition to the isolation variables, a method to perform an analytic reconstruction of the $\B\to\tautau$ decay chain, described in detail in Refs.~\cite{CERN-THESIS-2015-264, supp}, has been developed.
It combines geometrical information about the decay and mass constraints on the particles ($\B$, $\tau$ and $\nu$) in the decay chain to calculate the $\tau$ momenta analytically.
The possible solutions for the two $\tau$ momenta are found as solutions of a system of two coupled equations of second degree with two unknowns.
The finite detector resolution and approximations made in the calculation prevent real solutions being found for a substantial fraction of the signal events.
However, several intermediate quantities associated with the method are exploited to discriminate signal from background.

To make full use of the discrimination power present in the distributions of the selection variables, a requirement is added on the output of a neural network \cite{Feindt:2004wla}, built using seven variables: the $\tau^{\pm}$ candidate masses and decay times, a charged track isolation variable for the pions, a neutral isolation variable for the $B$ candidate, and one variable from the analytic reconstruction method, introduced in Ref.~\cite{supp}.
The classifier is trained on simulated $\B\to\tautau$ decays, representing the signal, and data events from the signal-depleted region.

In order to determine the signal yield, a binned maximum likelihood fit is performed on the output of a second neural network (NN), built with 29 variables and using the same training samples.
The NN inputs include the eight variables from the analytic reconstruction method listed in Ref.~\cite{supp}, further isolation variables, as well as kinematic and geometrical variables.
The NN output is transformed to obtain a flat distribution for the signal over the range [0.0, 1.0], while the background peaks towards zero.

Varying the two-pion invariant mass sector boundaries, the signal region is optimised for the $\Bs\to\tautau$ branching fraction limit using pseudoexperiments.
The boundaries are set to 615 and 935\,\mevcc.
The overall efficiency of the selection, determined using simulated $\B_{(s)}^0\to\tautau$ decays, is approximately $2.2(2.4)\times 10^{-5}$, including the geometrical acceptance.
Assuming the SM prediction, the number of $\Bs\to\tautau$ decays expected in the signal region is 0.02.

\begin{figure}[!t]
\center
\includegraphics[width=0.49\textwidth]{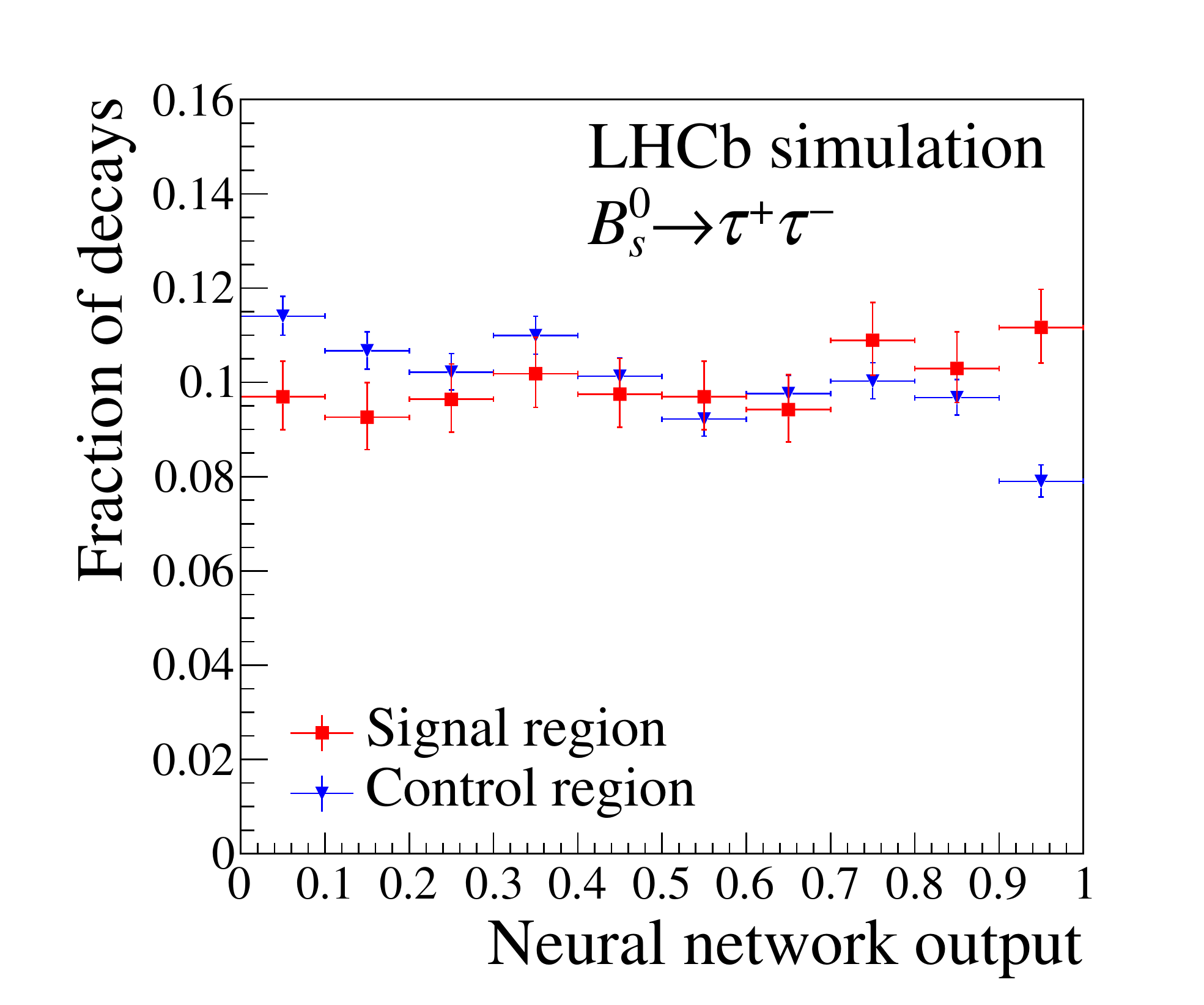} 
\hfill
\includegraphics[width=0.49\textwidth]{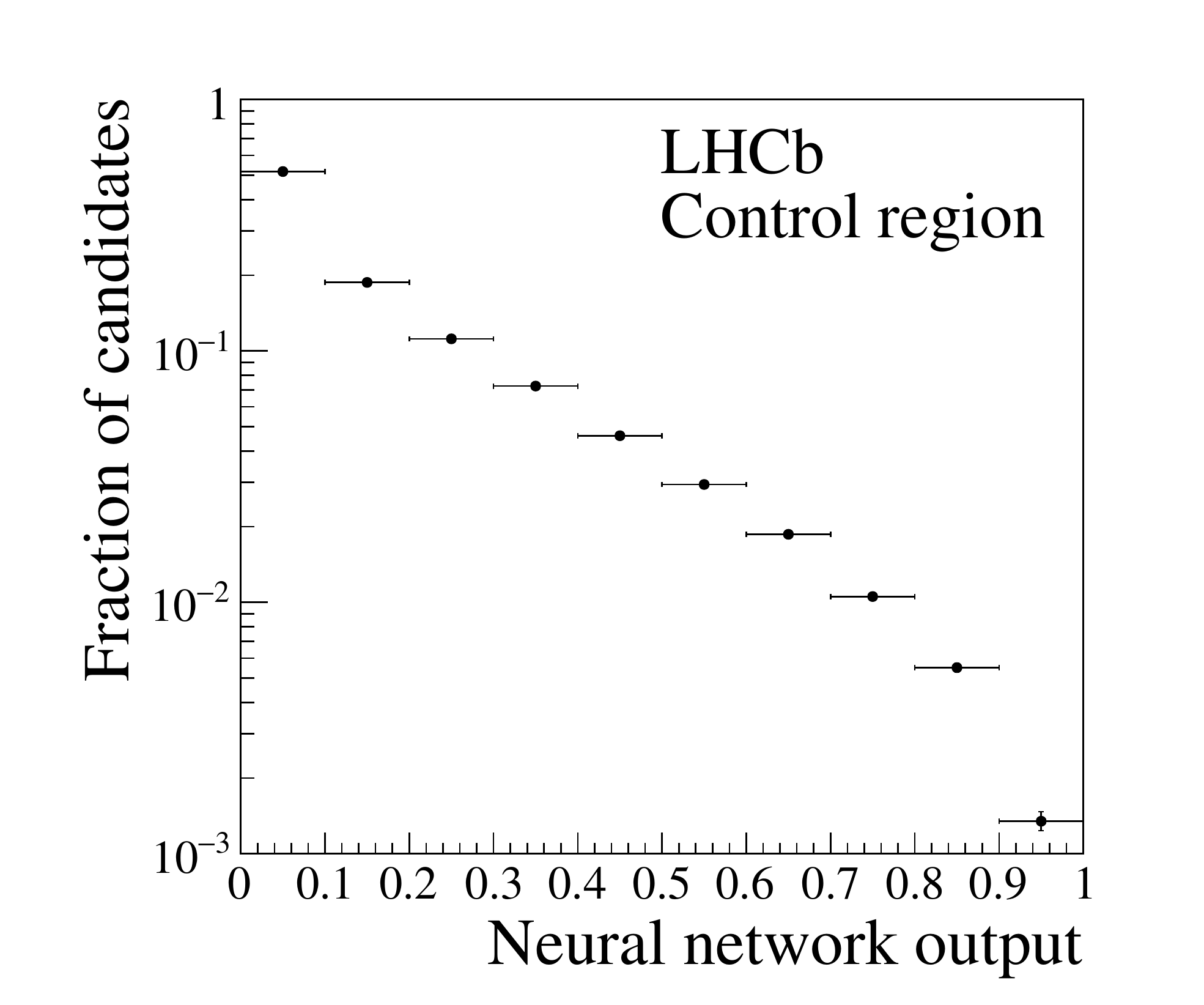}
\caption{(left) Normalised NN output distribution in the signal ($\widehat{\mathcal{N}}^{\text{SR}}_{\text{sim}}$) and control ($\widehat{\mathcal{N}}^{\text{CR}}_{\text{sim}}$) region for $\Bs\to\tautau$ simulated events. (right) Normalised NN output distribution in the data control region $\widehat{\mathcal{N}}^{\text{CR}}_{\text{data}}$.
The uncertainties reflect the statistics of the (simulated) data.}
\label{Fig:Templates}
\end{figure}

%%%%%%%%%%%%%%%%%%%%%%%%%%%%%%%%%%%%%%%%%%%%%%%%%%
%%% Fit
After the selection, the signal, signal-depleted and control regions contain, respectively, 16\%, 13\% and 58\% of the simulated signal decays.
The corresponding fractions of selected candidates in data are 7\%, 37\% and 47\%.
Most signal decays fall into the control region, but the signal region, which contains about 14\,700 candidates in data after the full selection, is more sensitive due to its lower background contamination.
For the fit, ten equally sized bins of NN output in the range [0.0, 1.0] are considered, where the high NN region [0.7,~1.0] was not investigated until the fit strategy was fixed.
The signal model is taken from the $\Bs\to\tautau$ simulation, while the background model is taken from the data control region, correcting for the presence of expected signal events in this region.
The fit model is given by
\begin{equation}\label{Eq:fit}
\mathcal{N}^{\text{SR}}_{\text{data}} = s\times \widehat{\mathcal{N}}^{\text{SR}}_{\text{sim}} + f_b\times\left(\mathcal{N}^{\text{CR}}_{\text{data}} - s \times\frac{\eps^{\text{CR}}}{\eps^{\text{SR}}}\times\widehat{\mathcal{N}}^{\text{CR}}_{\text{sim}}\right)\,, 
\end{equation}
where $\mathcal{N}^{\text{SR}}_{\text{sim/data}}$ ($\mathcal{N}^{\text{CR}}_{\text{sim/data}}$) is the NN output distribution in the signal (control) region from simulation/data, $s$ is the signal yield in the signal region, $f_b$ is a scaling factor for the background template, and $\eps^{\text{SR}}$ ($\eps^{\text{CR}}$) is the signal efficiency in the signal (control) region.
The quantities $s$ and $f_b$ are left free in the fit.
The corresponding normalised distributions $\widehat{\mathcal{N}}^{\text{SR}}_{\text{sim}}$, $\widehat{\mathcal{N}}^{\text{CR}}_{\text{sim}}$ and $\widehat{\mathcal{N}}^{\text{CR}}_{\text{data}}$ are shown in Fig.~\ref{Fig:Templates}.
%%%%%%%%%%%%%%%%%%%%%
%%%%%%%%%%%%%%%%%%%%

The agreement between the background NN output distributions in the control and signal regions has been tested in different samples: in the data for the background-dominated NN output bins [0.0, 0.7], in a generic $b\overline{b}$ simulated sample and in several specific simulated background modes (such as $\Bd\to\Dm\pip\pim\pip$ with $\Dm\to\Kz\pim\pip\pim$, or $\Bs\to\Dsm\pip\pim\pip$ with $\Dsm\to\taum\nu_{\tau}$).
Within the statistical uncertainty, the distributions have been found to agree with each other in all cases.
The background in the control region can therefore be used to characterise the background in the signal region.

Differences between the shapes of the background distribution in the signal and control regions of the data are the main sources of systematic uncertainties on the background model.
These uncertainties are taken into account by allowing each bin in the $\mathcal{N}^{\text{CR}}_{\text{data}}$ distribution  to vary according to a Gaussian constraint. 
The width of this Gaussian function is determined by splitting the control region into two approximately equally populated samples and taking, for each bin, the maximum difference between the NN outputs of the two subregions and the unsplit sample.
The splitting is constructed to have one region more signal-like and one region more background-like.

\begin{figure}
\center
\includegraphics[width=0.7\textwidth]{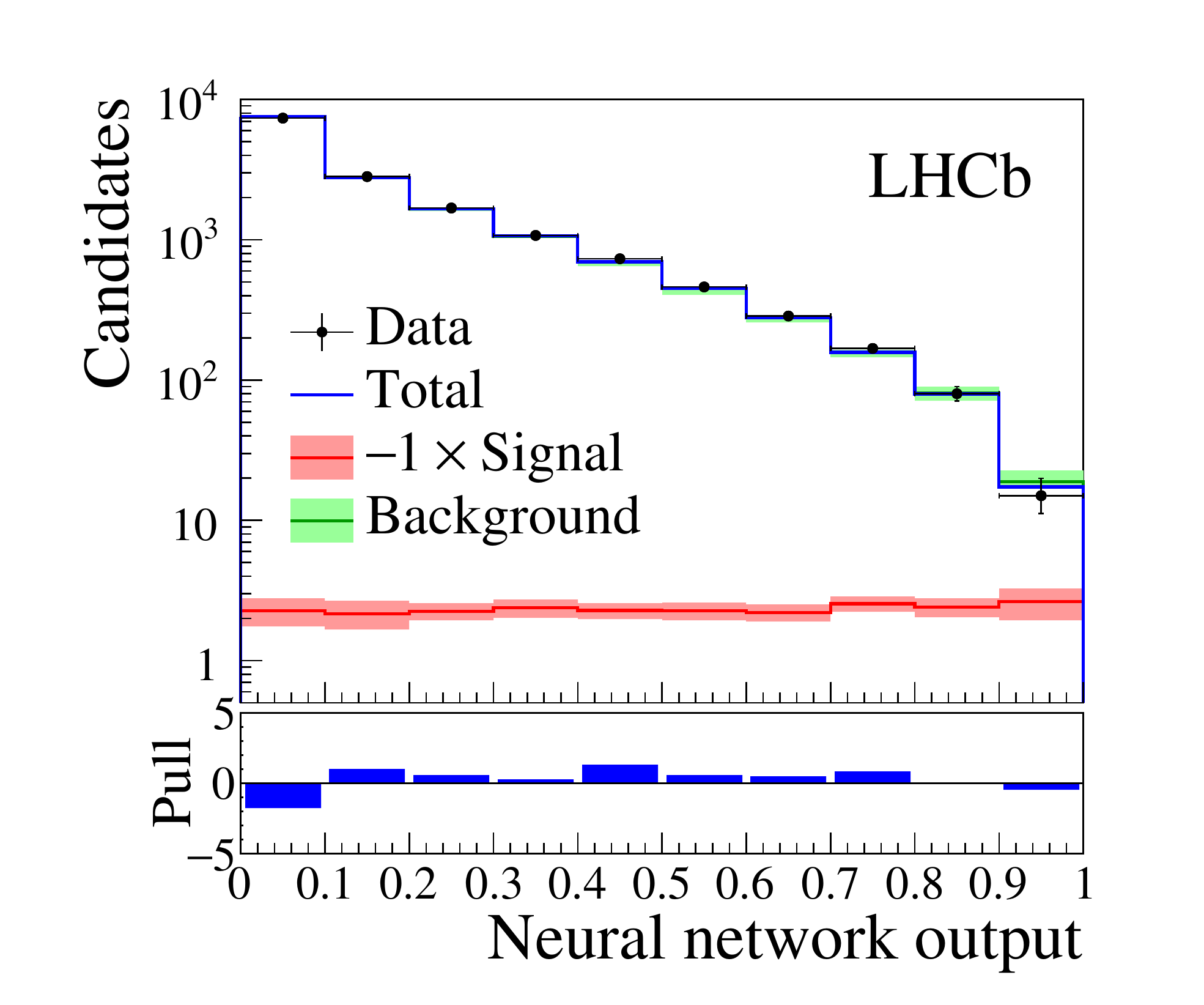}
\caption{Distribution of the NN output in the signal region $\mathcal{N}^{\text{SR}}_{\text{data}}$ (black points), with the total fit result (blue line) and the background component (green line).
The fitted $\Bs\to\tautau$ signal component is negative and is therefore shown multiplied by $-1$ (red line).
For each bin of the signal and background component the combined statistical and systematic uncertainty on the template is shown as a light-coloured band.
The difference between data and fit divided by its uncertainty (pull) is shown underneath.}
\label{Fig:NN_output_Unblind_Strategy_3}
\end{figure}

The signal can be mismodelled in the simulation.
The $\Bd\to\Dm\Dsp$ decay is used to compare data and simulation for the variables used in the NN.
Ten variables are found to be slightly mismodelled and their distributions are corrected by weighting.
The difference in the shape of the NN output distribution compared to the original unweighted sample is used to derive the associated systematic uncertainty.
The fit procedure is validated with pseudoexperiments and is found to be unbiased.
Assuming no signal contribution, the expected statistical (systematic) uncertainty on the signal yield is $^{+62}_{-40}\:(^{+40}_{-42})$.
The fit result on data is shown in Fig.~\ref{Fig:NN_output_Unblind_Strategy_3} and gives a signal yield $s = -23^{+63}_{-53} \stat ^{+41}_{-40} \syst$, where the split between the statistical and systematic uncertainties is based on the ratio expected from pseudoexperiments.

%%%%%%%%%%%%%%%%%%%%%
%%%%%%%%%%%%%%%%%%%%

\begin{figure}
\center
\includegraphics[width=0.7\textwidth]{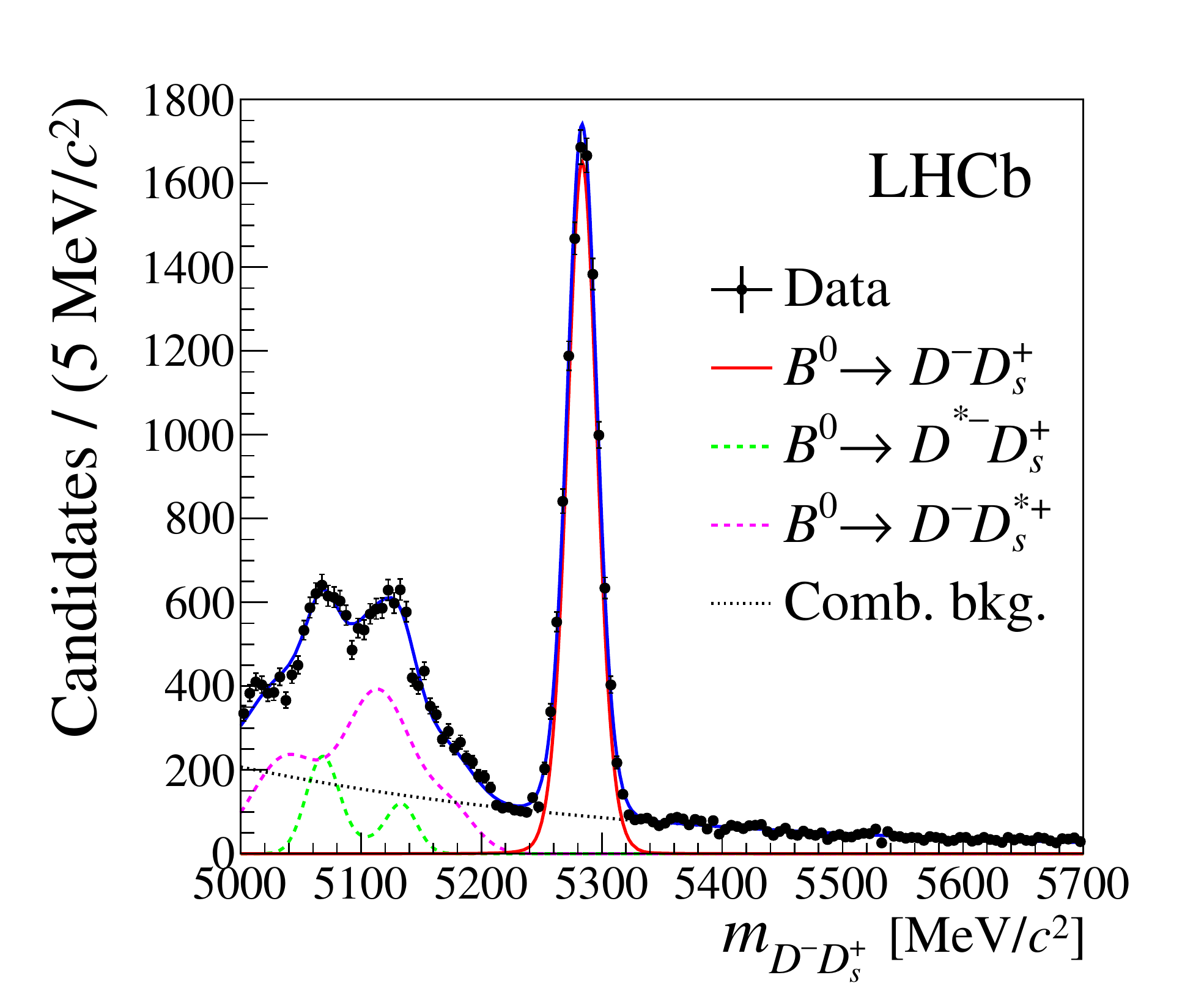}
\caption{Invariant mass distribution of the reconstructed $\Bd\to D^-D_s^{+}$ candidates in data (black points), together with the total fit result (blue line) used to determine the $\Bd\to D^-D_s^{+}$ yield.
The individual components are described in the text.}
\label{Fig:B2DD_massfit}
\end{figure}

%%%%%%%%%%%%%%%%%%%%%%%%%%%%%%%%%%%%%%%%%%%%%%%%%%
%%% Normalisation
The $\Bs\to\tautau$ signal yield is converted into a branching fraction using \mbox{$\BR(\Bs\to\tautau)$} $=\alpha^s\times s$, with
\begin{equation}\label{Eq:Alpha_norm}
\alpha^s \equiv \frac{\epsilon^{D^-D_s^+} \times \BR(\Bd\to\Dm\Dsp)\times\BR(\Dp\to \Km\pip\pip)\times\BR(\Dsp\to \Kp\Km\pip)}{N_{\Dm\Dsp}^{\text{obs}}\times\epsilon^{\tautau}\times \left[\BR(\tau^-\to\pi^-\pi^+\pi^-\nu_{\tau})\right]^2}\times\frac{f_d}{f_s}\:,
\end{equation}
where $\epsilon^{\tautau}$ and $\epsilon^{D^-D_s^+}$ are the combined efficiencies of trigger, reconstruction and selection of the signal and normalisation channels.
The branching fractions used are \mbox{$\BR(\Bd\to\Dm\Ds) = (7.5 \pm 1.1)\times 10^{-3}$} \cite{Zupanc:2007pu}, \mbox{$\BR(\Dm\to\Kp\pim\pim) = (9.46 \pm 0.24)\%$} \cite{PDG2016} and \mbox{$\BR(\Dsp\to\Km\Kp\pip) = (5.45 \pm 0.17)\%$} \cite{PDG2016}, and $f_s/f_d = 0.259 \pm 0.015$ \cite{LHCb-PAPER-2012-037, *LHCb-CONF-2013-011} is the ratio of \Bs to \Bd production fractions.
The efficiencies are determined using simulation, applying correction factors derived from data.
The \Bd\to\Dm\Dsp yield, $N_{D^-D_s^+}^{\text{obs}}$, is obtained from a fit to the  mass distribution, which has four contributions: 
the \Bd\to\Dm\Dsp component, modelled by a Hypatia function~\cite{Santos:2013gra}, a combinatorial background component, described by an exponential function, and two partially reconstructed backgrounds, $\Bd\to D^{*-}\Dsp$ and $\Bd\to \Dm D_s^{*+}$, modelled as in Ref.~\cite{LHCb-PAPER-2012-050}.
The resulting fit is shown in Fig.~\ref{Fig:B2DD_massfit} and gives a yield of $N_{D^-D_s^+}^{\text{obs}} = 10\,629 \pm 114$, where the uncertainty is statistical.

Uncertainties on $\alpha^s$ arise from the $\Bd\to D^-D_s^{+}$ fit model, the finite size of the simulated samples, the uncertainty from the corrections to the simulation and external inputs.
The latter contribution, which includes the branching fractions and hadronisation fractions in Eq.~\eqref{Eq:Alpha_norm}, is dominant, giving a relative uncertainty of 17\% on $\alpha^s$.
The $\Bd\to D^-D_s^{+}$ fit model is varied using the sum of two Gaussian functions with a common mean and power-law tails instead of the Hypatia function for the signal, a second-order Chebychev polynomial instead of an exponential function for the combinatorial background, and adding two other background components from $\Bs\to D^-D_s^{*+}$ and $\Bd\to a_1(1260)^-D_s^{*+}$ decays.
The change in signal yield compared to the nominal fit is taken as a systematic uncertainty, adding the contributions from the four variations in quadrature.
The overall relative uncertainty on $\alpha^s$ due to $N_{D^-D_s^+}^{\text{obs}}$ (including the fit uncertainty) is 1.7\%.
Corrections determined from $J/\psi\to\mu^+\mu^-$ and $D^0\to K^-\pi^+$ data control samples are applied for the tracking, PID and the hadronic hardware trigger efficiencies.
The relative uncertainty on $\alpha^s$ due to selection efficiencies is 2.9\%, taking into account both the limited size of the simulated samples and the systematic uncertainties.
The normalisation factor is found to be $\alpha^s= (4.07 \pm 0.70)\times10^{-5}$.

The shapes of the NN output distributions and the selection efficiencies depend on the parametrisation used in the simulation to model the $\taum\to\pim\pip\pim\nu_{\tau}$ decay.
The result obtained with the \textsc{Tauola} \babar-tune model is therefore compared to available alternatives \cite{Was:2015laa}, which are based on CLEO data for the $\taum\to\pim\pi^0\pi^0\nu_{\tau}$ decay \cite{Asner:1999kj}.
The selection efficiency for these alternative models can be up to 20\% higher, due to different structures in the two-pion invariant mass, resulting in lower limits.
Dependence of the NN signal output distribution on the $\tau$-decay model is found to be negligible.
Since the alternative models are based on a different $\tau$ decay, the \babar-tune model is chosen as default and no systematic uncertainty is assigned.

%%%%%%%%%%%%%%%%%%%% 
%%%%%%%%%%%%%%%%%%%% 

%%%%%%%%%%%%%%%%%%%%%%%%%%%%%%%%%%%%%%%%%%%%%%%%%%
%%% Results
The signal yield obtained from the likelihood fit is translated into an upper limit on the $\Bs\to\tautau$ branching fraction using the CL$_{\text{s}}$ method \cite{Read:2002hq,Cowan:2010js}.
Assuming no contribution from $\Bd\to\tautau$ decays, an upper limit is set on the $\Bs\to\tautau$ branching fraction of $ 5.2\,(6.8) \times 10^{-3}$ at 90\,(95)\% CL.
This is the first experimental limit on $\BR(\Bs\to\tautau)$.
The analysis is repeated for the $\Bd\to\tautau$ decay.
The fit is performed by replacing the signal model with that derived from simulated $\Bd\to\tautau$ decays, giving $s = -15^{+67}_{-56} \stat ^{+44}_{-42} \syst$ \cite{supp}.
The expected statistical (systematic) uncertainty on the signal yield is $^{+64}_{-58}\:(^{+41}_{-43})$.
The corresponding normalisation factor is $\alpha^d = (1.16 \pm 0.19)\times10^{-5}$.
The limit obtained is $\BR(\Bd\to\tautau)< 1.6\,(2.1) \times 10^{-3}$ at 90\,(95)\% CL, which constitutes a factor 2.6 improvement with respect to the \babar result \cite{Aubert:2005qw} and is the current best limit on $\BR(\Bd\to\tautau)$.

%%%%% End Matter
%!TEX root = main.tex

%%%%%%%%%%%%%%%%%%%%%%%%%%%%%%%%%%%%%%%%%%%%%%%%%%
\section*{Acknowledgements}
 
\noindent We thank J\'er\^ome Charles (CPT, Marseille, France) for fruitful discussions and help in developing the analytic reconstruction method.
%%% Official LHCb file
We express our gratitude to our colleagues in the CERN
accelerator departments for the excellent performance of the LHC. We
thank the technical and administrative staff at the LHCb
institutes. We acknowledge support from CERN and from the national
agencies: CAPES, CNPq, FAPERJ and FINEP (Brazil); MOST and NSFC (China);
CNRS/IN2P3 (France); BMBF, DFG and MPG (Germany); INFN (Italy); 
FOM and NWO (The Netherlands); MNiSW and NCN (Poland); MEN/IFA (Romania); 
MinES and FASO (Russia); MinECo (Spain); SNSF and SER (Switzerland); 
NASU (Ukraine); STFC (United Kingdom); NSF (USA).
We acknowledge the computing resources that are provided by CERN, IN2P3 (France), KIT and DESY (Germany), INFN (Italy), SURF (The Netherlands), PIC (Spain), GridPP (United Kingdom), RRCKI and Yandex LLC (Russia), CSCS (Switzerland), IFIN-HH (Romania), CBPF (Brazil), PL-GRID (Poland) and OSC (USA). We are indebted to the communities behind the multiple open 
source software packages on which we depend.
Individual groups or members have received support from AvH Foundation (Germany),
EPLANET, Marie Sk\l{}odowska-Curie Actions and ERC (European Union), 
Conseil G\'{e}n\'{e}ral de Haute-Savoie, Labex ENIGMASS and OCEVU, 
R\'{e}gion Auvergne (France), RFBR and Yandex LLC (Russia), GVA, XuntaGal and GENCAT (Spain), Herchel Smith Fund, The Royal Society, Royal Commission for the Exhibition of 1851 and the Leverhulme Trust (United Kingdom).

\addcontentsline{toc}{section}{References}
\setboolean{inbibliography}{true}
\bibliographystyle{LHCb}
\bibliography{tautau}

%%%%% Supplemental Material
\clearpage
\appendix
%!TEX root = main.tex

%%%%%%%%%%%%%%%%%%%%%%%%%%%%%%%%%%%%%%%%%%%%%%%%%%
\section{Supplemental material}\label{sec:Supplementary-App}

In Section A.1, details about the analytic reconstruction method are given. Sections A.2 and A.3 contain additional results related to the $\Bs\to\tautau$ and $\Bd\to\tautau$ decay channels.

%%%%%%%%%%%%%%%%%%%%%%%%%%%%%%%%%%%%%%%%%%%%%%%%%%
\subsection{Reconstruction method}\label{sec:reconstruction}
A method to perform an analytic reconstruction of the $\B\to\tautau$ decay chain is described in the following and in detail in Ref.~\cite{CERN-THESIS-2015-264}.
It combines geometrical information about the decay and sets mass constraints on the particles in the decay chain ($\Bs$, $\tau$ and $\nu$) to calculate the $\tau$ momenta analytically.
In these calculations, Lorentz invariance is kept manifest and the possible values for the two $\tau$ momenta are found as analytic solutions of a system of two coupled equations of second degree in two unknowns.
The only remaining degree of freedom is a single Lorentz scalar, introduced below as the angle $\theta$, measuring the asymmetry of the decay triangle in the decay-time space of the two $\tau$ leptons.

In the following, the unknown momenta of the $\tau^{\pm}$ leptons, the primary parameters of interest, are labelled by the four-vectors $p_{\pm}^{\mu}$;
the $\Bs\to\tautau$ decay plane is defined by the $B$ production vertex, \ie the PV, and the two $\tau$ decay vertices;
the three-vectors pointing from the PV to the $\tau^{\pm}$ decay vertices are labelled $\vec{w}_{\pm}$;
the time intervals between the $B$ production and the $\tau^{\pm}$ decays, $w^0_{\pm}$, which cannot be measured, act as the temporal counterpart to $\vec{w}_{\pm}$.
Together, they make the four-vectors $w_{\pm}^{\mu} = (w^0_{\pm}, \vec{w}_{\pm})$.
Introducing the notation
\begin{equation}
W\equiv (w_+^{\mu}, w_-^{\mu})\:,\qquad P\equiv (p_+^{\mu}, p_-^{\mu})\:,
\end{equation}
the momenta $p_{\pm}^{\mu}$ can now be obtained from the coupled set of equations
\begin{equation}\label{Eq:WHP}
W = H\cdot P
\end{equation}
by imposing momentum conservation throughout the decay chain.
Here
\begin{equation}
H \equiv \left(
\begin{array}{cc}
\frac{t_B}{m_B} + \frac{t_{\tau_+}}{m_\tau} & \frac{t_B}{m_B}\\
\frac{t_B}{m_B} & \frac{t_B}{m_B} + \frac{t_{\tau_-}}{m_\tau}
\end{array}
\right)
\end{equation}
is given in terms of the $B$ and $\tau$ masses, $m_i$, and decay times, $t_i$.
Using the on-shell and flight-direction constraints, it is possible to rewrite Eq.~\eqref{Eq:WHP} in terms of a single unknown, chosen to be the rotation angle, $\theta$, that diagonalises the matrix $H$.
Thanks to this transformation, solving Eq.~\eqref{Eq:WHP} becomes equivalent to finding the roots of a fourth-order polynomial
\begin{equation}\label{Eq:TauReco_Poly}
\mathcal{P}^{(4)}(\xi)=\sum_{i=0}^4 a_i(\theta)\xi^i=0\:,
\end{equation}
where explicit expressions for the coefficients $a_i(\theta)$ in terms of $\theta$ and measurable quantities are given in Appendix D of Ref.~\cite{CERN-THESIS-2015-264}.

Even though it is possible, in principle, to exactly determine the angle $\theta$, a different approach has been used, because of the high complexity of the trigonometric equations involved. 
The value of the angle $\theta$ is in fact approximated in the calculation of the complex solutions $\xi_1$, \ldots, $\xi_4$.
Three approximations have been considered in Ref.~\cite{CERN-THESIS-2015-264}.
They are
\begin{enumerate}
\item $\theta = \langle\theta\rangle = \pi/4$, representing the case where both $\tau$ leptons have the same decay time.
\item $\theta = \bar{\theta}$, where $\bar{\theta}$ is the angle diagonalising the matrix
\begin{equation}
\overline{H} \equiv \left(
\begin{array}{cc}
1 + \frac{|\vec{w}_+|}{|\vec{w}_+ + \vec{w}_-|} & 1\\
1 & 1 + \frac{|\vec{w}_-|}{|\vec{w}_+ + \vec{w}_-|}
\end{array}
\right)
\end{equation}
\item $\theta = \theta^*$, where $\theta^*$ is the result of applying iterative corrections to $\theta = \bar{\theta}$.
\end{enumerate}

These approximations, together with the finite detector resolution, prevent having real solutions for a substantial fraction of the signal events.
Nonetheless, quantities appearing in intermediate calculations, though not having immediate physical meaning, have been found useful to discriminate between signal and background.
The most powerful of these variables are exploited in the two neural networks that are used in the candidate selection.
The variable
\begin{itemize}
  \item $\Real[|\vec{\tilde{p}}_{+}(\theta=\bar{\theta},\xi_3)|]$
\end{itemize}
is used by both the first and second NN, while the seven variables
\begin{itemize}                                                                            
  \item $\Imag[|\vec{\tilde{p}}_{+}(\theta=\theta^\star,\xi_1)|]$,   
  \item $\Imag[\tilde{p}_{+}(\theta=\theta^\star,\xi_1)\tilde{p}_{-}(\theta=\theta^\star,\xi_1)]$,                                       
  \item $\Imag[\tilde{p}_{+}(\theta=\theta^\star,\xi_3)\tilde{p}_{-}(\theta=\theta^\star,\xi_3)]$,                                              
  \item $\Real[\tilde{p}_{+}(\theta=\pi/4,\xi_1)\tilde{p}_{-}(\theta=\pi/4,\xi_1)]$,  
  \item $\Real[\xi_1(\theta=\theta^*)]$,                                                                   
  \item $\bar{\theta}$,
  \item $p_+\cdot p_-$,
\end{itemize}
are used by the second NN.
Here $\tilde{p}^2_{\pm}\equiv M^2_{\pm}=m^2_{\tau}\pm\sin(2\theta)\hat{s}$, and $\hat{s}\equiv p_+\cdot p_-$ (for signal events $\hat{s}=\frac{M^2_B-2m^2_{\tau}}{2}$).

\clearpage
%%%%%%%%%%%%%%%%%%%%%%%%%%%%%%%%%%%%%%%%%%%%%%%%%%
\subsection{Additional $\Bs$ fit result}\label{sec:add_bs}

\begin{figure}[bp]
\center
\includegraphics[width=0.6\textwidth]{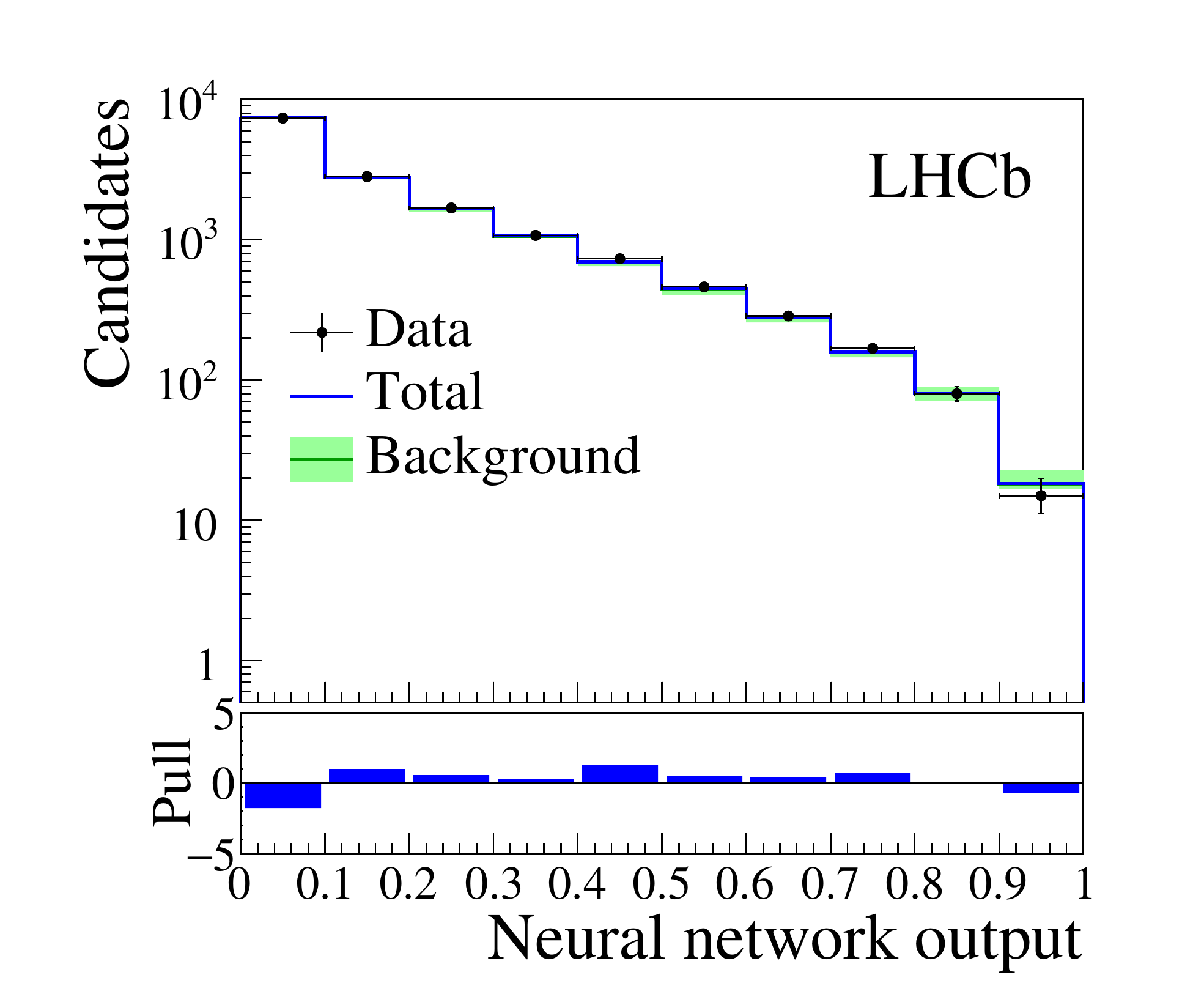}
\caption{Distribution of the NN output in the signal region  $\mathcal{N}^{\text{SR}}_{\text{data}}$ (black points), with the total fit result (blue line), and the background component (green line).
Shown is the fit using the ``background only'' model.
For each bin of the background component the combined statistical and systematic uncertainty is shown as a light-coloured band.
The difference between data and fit divided by its uncertainty (pull) is shown underneath.}
\label{Fig:NN_output_Unblind_Strategy_3_bkg}
\end{figure}

\begin{figure}[tbp]
\center
\includegraphics[width=0.6\textwidth]{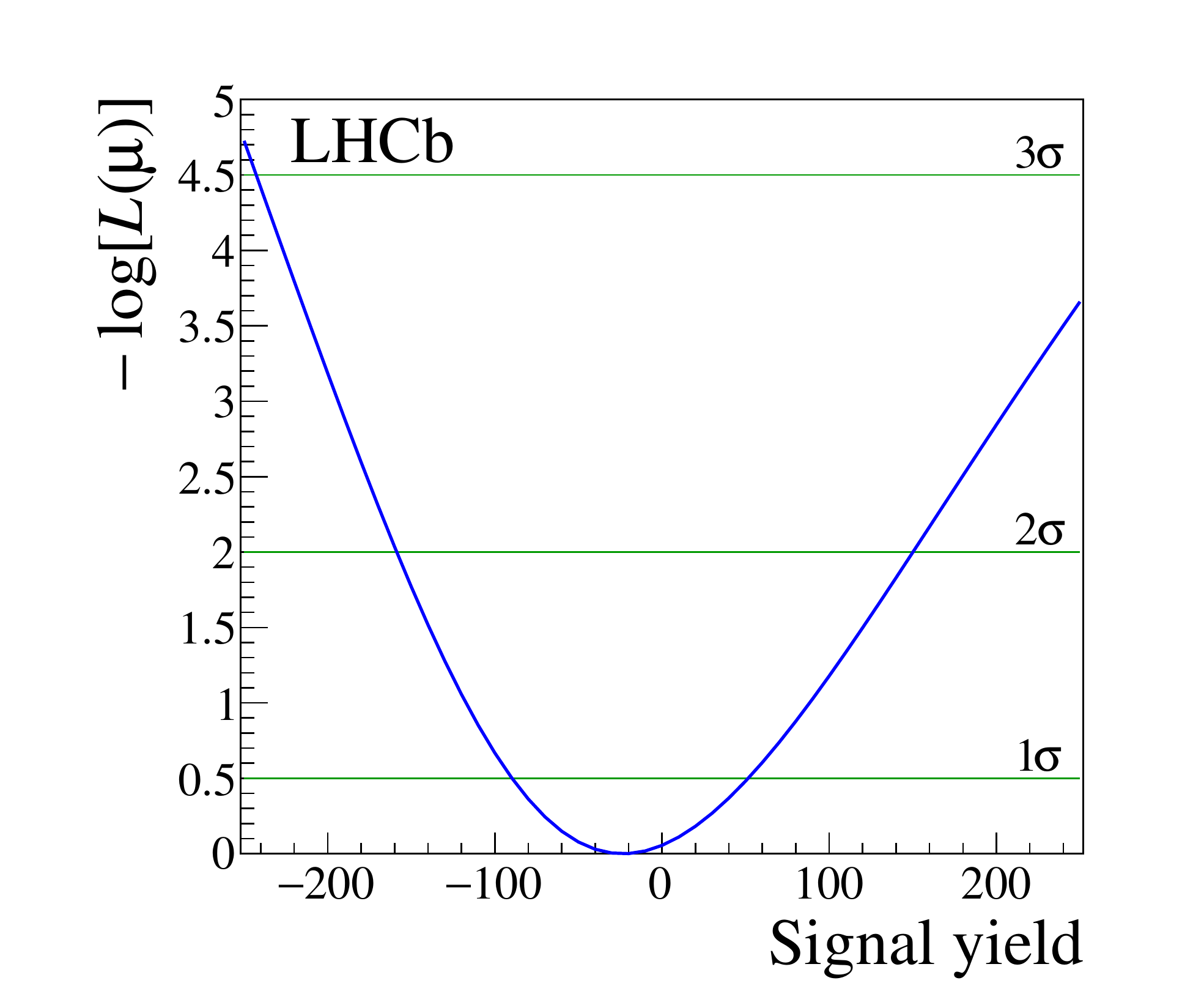}
\caption{Profile likelihood scan of the $\Bs\to\tautau$ signal yield.
The intersections of the likelihood curve with the horizontal lines define the 68.3\%, 95.4\% and 99.7\% likelihood intervals.}
\label{Fig:NLL_Scan_Unblind_Strategy_3}
\end{figure}

\begin{figure}[tbp]
\center
\includegraphics[width=0.7\textwidth]{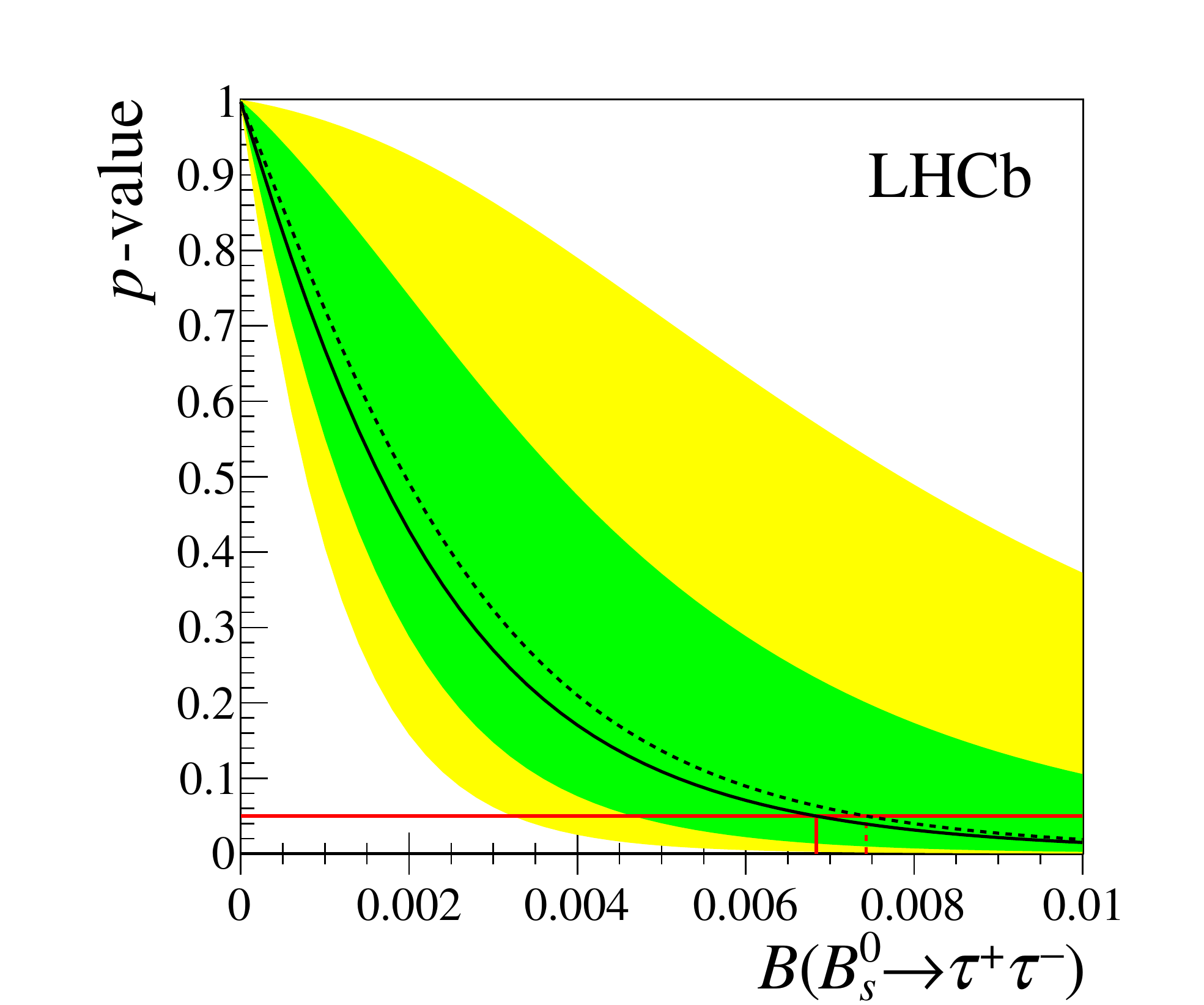}
\caption{The $p$-value derived with the CL$_{\text{s}}$ method as a function of $\BR(\Bs\to\tautau)$. Expected (observed) values are shown by a dashed (full) black line.
The green (yellow) band covers the regions of 68\% and 95\% confidence for the expected limit.
The red horizontal line corresponds to the limit at 95\% CL.}
\label{Fig:Unblind_CLs_Strategy_3}
\end{figure}

Figure \ref{Fig:NN_output_Unblind_Strategy_3_bkg} shows the fit result using only the background model.
A likelihood-ratio test is performed comparing the nominal fit with the background-only alternative.
The $p$-value of the likelihood-ratio test is 0.06, and the associated $z$-score is 1.60, showing that the data are consistent with the background-only hypothesis.
Figure \ref{Fig:NLL_Scan_Unblind_Strategy_3} shows the profile likelihood of the nominal fit.
Figure \ref{Fig:Unblind_CLs_Strategy_3} shows the expected and observed CL$_{\text{s}}$ values as a function of the branching fraction.
The expected limit for the $\Bs$ mode is \mbox{$\BR(\Bs\to\tautau)<5.7\,(7.4) \times 10^{-3}$} at 90\,(95)\% CL.

%%%%%%%%%%%%%%%%%%%%%%%%%%%%%%%%%%%%%%%%%%%%%%%%%%
\subsection{Additional $\Bd$ fit result}\label{sec:add_bz}

\begin{figure}
\center
\includegraphics[width=0.6\textwidth]{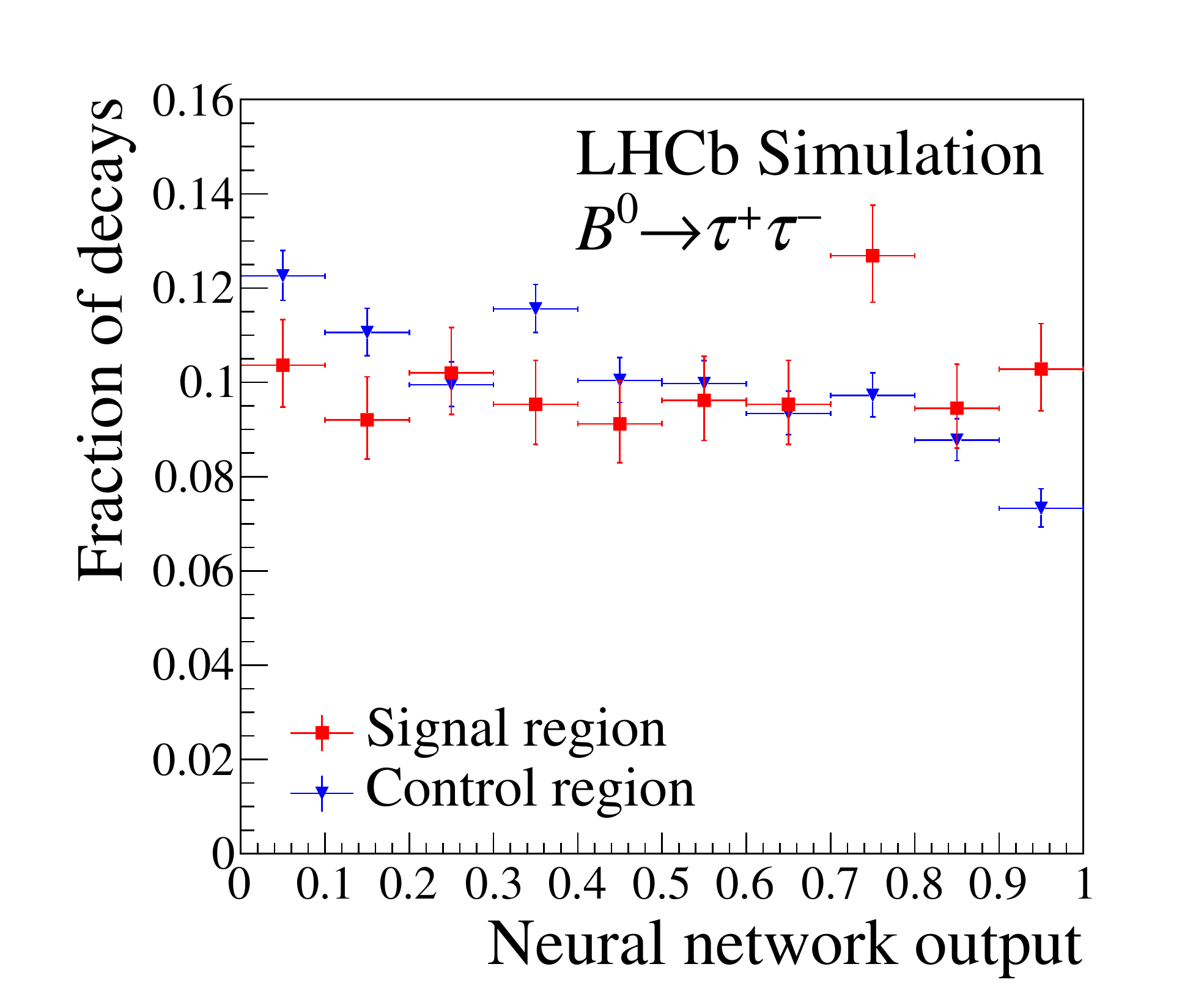}
\caption{Normalised NN output distribution in  the signal ($\widehat{\mathcal{N}}^{\text{SR}}_{\text{sim}}$) and control ($\widehat{\mathcal{N}}^{\text{CR}}_{\text{sim}}$) region for  $\Bz\to\tautau$ simulated events.}
\label{Fig:Template_Bd}
\end{figure}

\begin{figure}
\center
\includegraphics[width=0.6\textwidth]{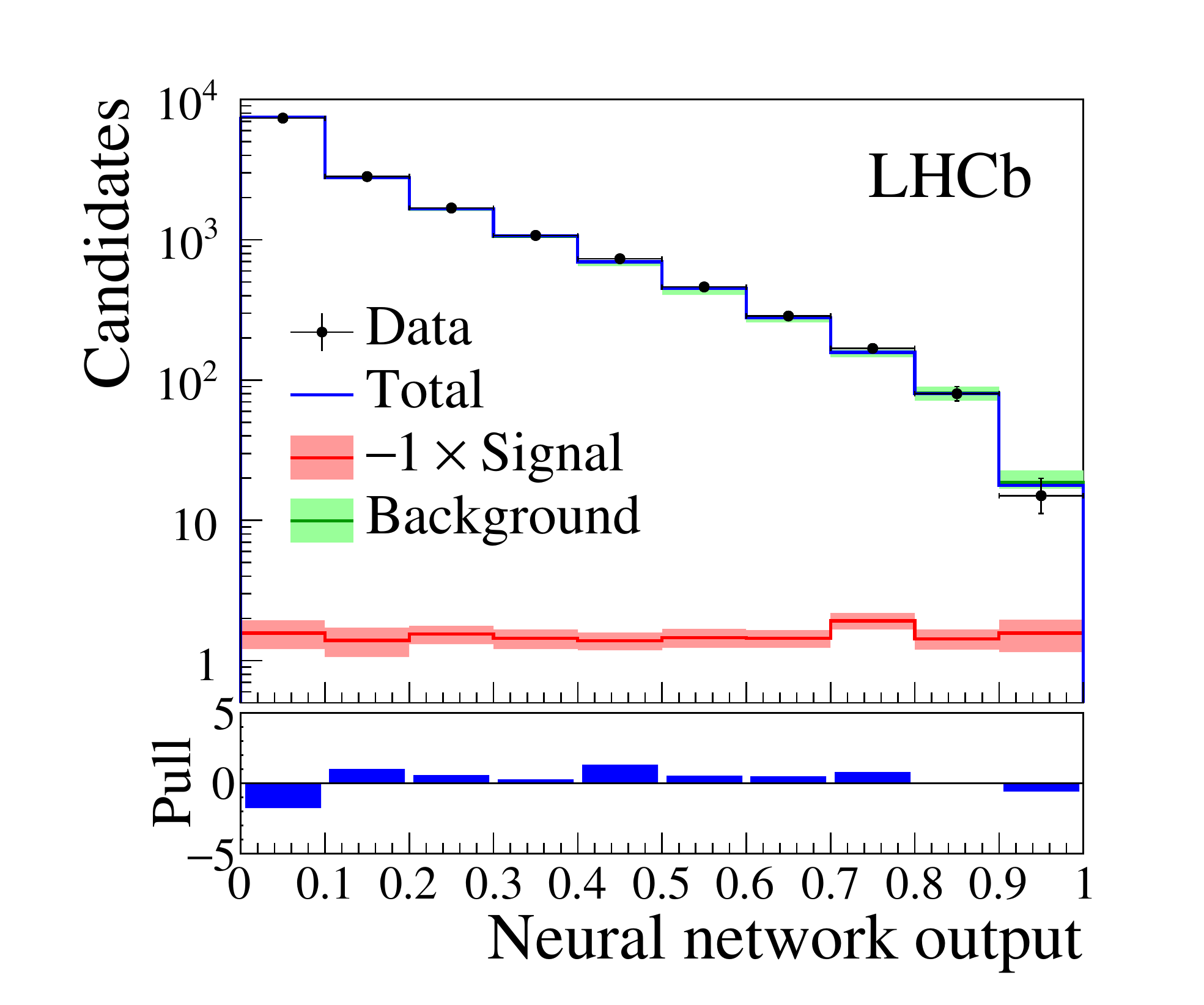}
\caption{Fit results for $\Bd\to\tautau$.
Distribution $\mathcal{N}^{\text{SR}}_{\text{data}}$  (black points), overlaid with the total fit result (blue line), and background component (green line), assuming all signal events originate from $\Bd\to\tautau$ decays.
The $\Bd\to\tautau$ signal component is also shown (red line), multiplied by $-1$.
For each bin of the signal and background component the combined statistical and systematic uncertainty is shown as a light-coloured band.
The difference between data and fit divided by its uncertainty (pull) is shown underneath.}
\label{Fig:Unblind_Fit_Bd}
\end{figure}

\begin{figure}
\center
\includegraphics[width=0.6\linewidth]{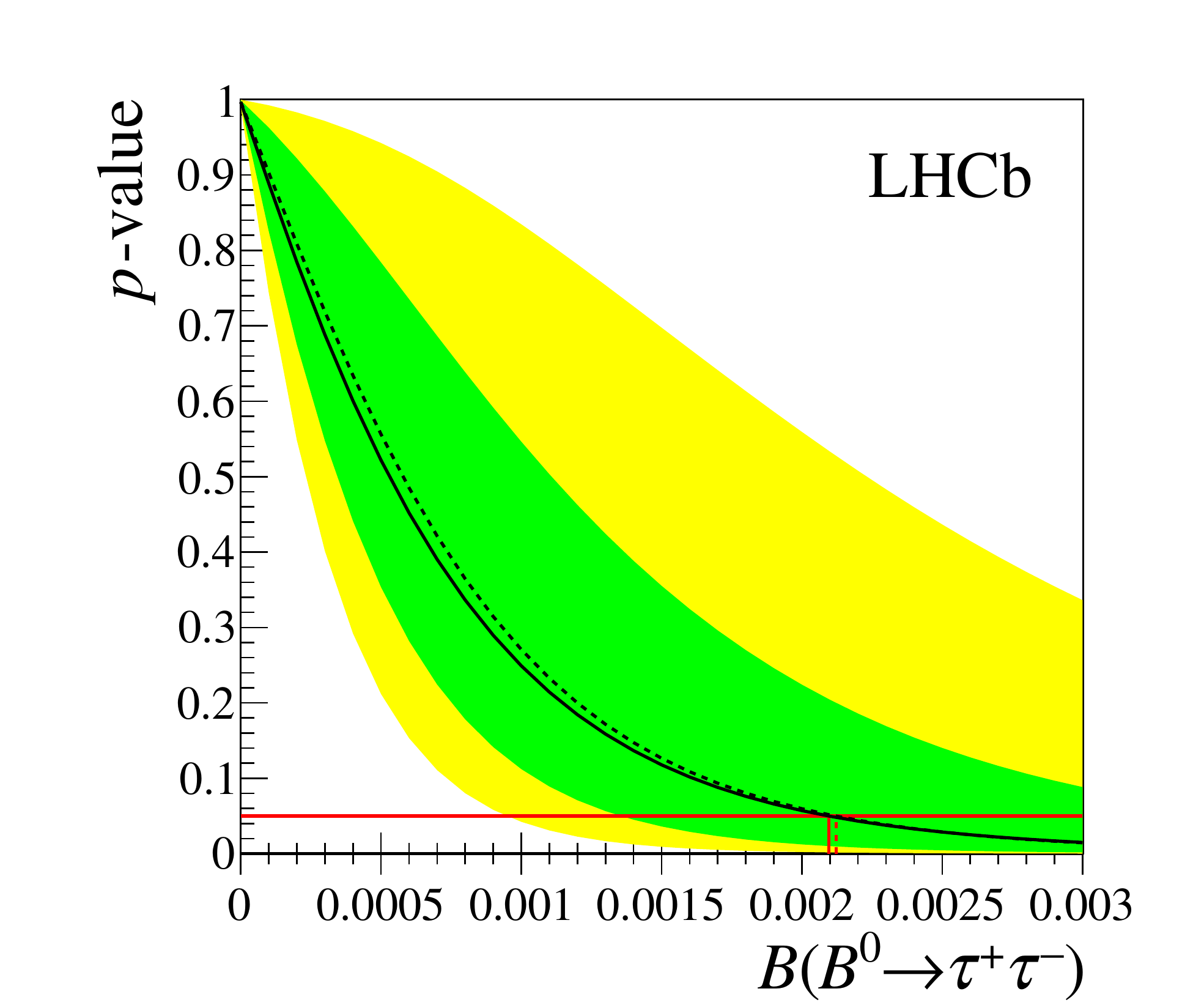}
\caption{The $p$-value derived with the CL$_{\text{s}}$ method as a function of  $\BR(\Bd\to\tautau$).
Expected (observed) values are shown by a dashed (full) black line.
The green (yellow) band covers the regions of 68\% and 95\% confidence for the expected limit.
The red horizontal line corresponds to the limit at 95\% CL.}
\label{Fig:Unblind_CLs_Bd}
\end{figure}

The NN output distributions for simulated $\Bd\to\tautau$ decays in the signal and control regions are shown in Fig.~\ref{Fig:Template_Bd}.
The fit result, assuming no contribution from \mbox{$\Bs\to\tautau$} decays, is shown in Fig.~\ref{Fig:Unblind_Fit_Bd}, and Fig.~\ref{Fig:Unblind_CLs_Bd} shows the expected and observed CL$_{\text{s}}$ values as a function of the branching fraction.
The expected limit for the  $\Bz$ mode is \mbox{$\BR(\Bz\to\tautau)<1.7\,(2.1) \times 10^{-3}$} at 90\,(95)\% CL.

%%%%%%%%%%%%%%%%%%%%%
%%%%%%%%%%%%%%%%%%%%%
%%%%%%%%%%%%%%%%%%%%%
%%%%%%%%%%%%%%%%%%%%%

%%%%%%%%%%%%%%%%%%%%
%%%%%%%%%%%%%%%%%%%%
%%%%%%%%%%%%%%%%%%%%
%%%%%%%%%%%%%%%%%%%% 

%%%%% Author List
\clearpage
\centerline{\large\bf LHCb collaboration}
\begin{flushleft}
\small
R.~Aaij$^{40}$,
B.~Adeva$^{39}$,
M.~Adinolfi$^{48}$,
Z.~Ajaltouni$^{5}$,
S.~Akar$^{59}$,
J.~Albrecht$^{10}$,
F.~Alessio$^{40}$,
M.~Alexander$^{53}$,
S.~Ali$^{43}$,
G.~Alkhazov$^{31}$
% ...

{\footnotesize \it
% ...
$ ^{5}$Clermont Universit{\'e}, Universit{\'e} Blaise Pascal, CNRS/IN2P3, LPC, Clermont-Ferrand, France\\
% ...
$ ^{10}$Fakult{\"a}t Physik, Technische Universit{\"a}t Dortmund, Dortmund, Germany\\
% ...
$ ^{31}$Petersburg Nuclear Physics Institute (PNPI), Gatchina, Russia\\
% ...
$ ^{39}$Universidad de Santiago de Compostela, Santiago de Compostela, Spain\\
$ ^{40}$European Organization for Nuclear Research (CERN), Geneva, Switzerland\\
% ...
$ ^{43}$Nikhef National Institute for Subatomic Physics, Amsterdam, The Netherlands\\
% ...
$ ^{48}$H.H. Wills Physics Laboratory, University of Bristol, Bristol, United Kingdom\\
% ...
$ ^{53}$School of Physics and Astronomy, University of Glasgow, Glasgow, United Kingdom\\
% ...
$ ^{59}$University of Cincinnati, Cincinnati, OH, United States\\
% ...
}
\end{flushleft}

\end{document}